\documentclass[11pt,a4paper]{article}
\usepackage{jheppub,xcolor}
\usepackage[utf8]{inputenc}
\usepackage{blindtext}
\usepackage{graphicx}
\usepackage{physics}
\usepackage{amsmath}
\usepackage{amsfonts}
\usepackage{amsthm}
\usepackage{amssymb}
\usepackage{caption}
\usepackage{booktabs}
\usepackage{subcaption}
\usepackage{slashed}
\usepackage{bigints}
\usepackage{setspace}
\usepackage{array}
\usepackage{comment}
\usepackage{multirow}
\usepackage{makecell} 
\definecolor{mygreen}{rgb}{0, 0.5019607843137255, 0}
\definecolor{Red}{rgb}{1.,0.,0.}
\definecolor{Blue}{rgb}{0.,0.,1.}
\newcommand{\CF}{C_F}
\newcommand{\CA}{C_A}
\allowdisplaybreaks

\newcommand{\ctG}{\frac{c_{tG}v}{\sqrt{2} \Lambda^2}}
\newcommand{\cG}{\frac{c_{G}}{\Lambda^2}}
\title{Two-loop renormalisation of quark and gluon fields in the SMEFT}

\author[a]{Claude Duhr,}
\author[a]{Andres Vasquez,}
\author[b]{Giuseppe Ventura,}
\author[b]{Eleni Vryonidou}

\emailAdd{cduhr@uni-bonn.de} \emailAdd{avasquez@uni-bonn.de} \emailAdd{giuseppe.ventura@manchester.ac.uk} \emailAdd{eleni.vryonidou@manchester.ac.uk}
\affiliation[a]{Bethe Center for Theoretical Physics, Universit\"{a}t Bonn, D-53115, Germany}
\affiliation[b]{Department of Physics and Astronomy, University of Manchester, Oxford Road, Manchester M13~9PL, United Kingdom}

\abstract{
We compute the contributions of CP-conserving operators in the dimension-six SMEFT to the two-loop renormalisation constants of quark and gluon fields in the $\overline{\textrm{MS}}$-scheme. Specifically, we consider the top-quark chromomagnetic operator and the triple gluon operator. We work with the background-field method, which allows us to extract the contribution of these operators to the two-loop running of the top mass and the strong coupling constant. We discuss in detail the mixing with the unphysical operators required for the renormalisation, and we present analytic results for the renormalisation constants of all relevant operators.
}

\begin{document}

\begin{flushright}
BONN-TH-2025-07
\end{flushright}

\maketitle

\section{Introduction}
The lack of evidence for new degrees of freedom in current collider experiments indicates a significant mass gap between the electroweak scale and the energy scale where new physics may emerge. Within this context, the Standard Model Effective Field Theory (SMEFT) \cite{Weinberg:1979sa, Leung:1984ni, Buchmuller:1985jz, 1008.4884} provides a powerful approach for exploring low-energy manifestations of high-scale new physics. By introducing higher-dimensional operators into the Lagrangian, the SMEFT captures short-distance deviations from the Standard Model (SM) predictions while imposing minimal assumptions on the underlying new physics.

Global SMEFT interpretations of data are ongoing \cite{2012.02779,Ethier:2021bye, Celada:2024mcf}, where precise theoretical predictions, combined with a plethora of experimental data, allow one to infer constraints on a large set of dimension-six Wilson coefficients. In this effort to constrain the scale of new physics, it has become clear that precise predictions in the effective field theory (EFT) can play a crucial role.  Higher-order predictions can enhance our sensitivity to new physics effects and will be vital in the interpretation of any confirmed deviation from the SM predictions. As such, efforts continue to systematically improve the SMEFT predictions. One-loop effects have been considered both in the Quantum Chromodynamics (QCD) and electroweak couplings. In the context of hadron collider physics, QCD corrections to SMEFT observables involving dimension-six operators have been automated in \cite{2008.11743}, whilst electroweak corrections have been computed on a process by process basis \cite{1911.11244,2201.09887,Hartmann:2015aia,Hartmann:2015oia,Gauld:2015lmb,Hartmann:2016pil,Dawson:2018dxp,Dawson:2018jlg,Dawson:2018liq,Dawson:2018pyl,Dedes:2018seb,Dedes:2019bew,Cullen:2019nnr,Boughezal:2019xpp,Cullen:2020zof,Corbett:2021cil,Dawson:2021ofa}. Two-loop computations for physical processes have also begun appearing in the literature, cf.,~e.g.,~\cite{1708.00460,1811.12366, Buchalla:2018yce,2202.02333,2204.13045, 2311.15004, 2410.13304,2409.05728,Braun:2025hvr}.

An important ingredient of precise EFT predictions is the inclusion of renormalisation group (RG) running of the SMEFT coefficients. The RG running controls the dependence of the Wilson coefficients on the energy scale through the corresponding anomalous dimension matrix, which has been computed at one-loop in \cite{1308.2627, 1310.4838, 1312.2014}. There are various publicly available computer codes to solve renormalisation group equations (RGEs) and to obtain the Wilson coefficients at different scales \cite{1704.04504, 2010.16341, 1804.05033, 1309.7030, 2210.06838}. RG effects have been shown to be significant for both SMEFT predictions and the subsequent extraction of the Wilson coefficients \cite{1607.05330, 2212.05067, 2109.02987, 2312.11327, 2406.06670, 2409.19578,2502.14030}. As a consequence, a precise knowledge of the RG running and the corresponding anomalous dimension matrix, including higher-order corrections in the SM couplings, is important to make reliable predictions for SMEFT observables. Beyond one-loop, increasing efforts have been made to determine the full form of the anomalous dimension matrix \cite{1910.05831,2005.12917,2308.06315, 2310.19883, 2311.13630,2211.09144, 2203.11224,2401.16904, 2408.03252, 2410.07320,2412.13251, 1907.04923, 2011.02494}. 

The anomalous dimension matrix governing the RGEs of the dimension-six SMEFT is directly connected to the ultraviolet (UV) structure of the theory. It is well known that higher-order computations may develop UV divergences that are removed during the process of renormalisation. A good understanding of the renormalisation of all relevant operators is a necessary condition for computing higher-loop corrections to a given SMEFT process. While the dimension-six SMEFT is not a renormalisable theory, it is renormalisable order by order in the EFT expansion, i.e., if all observables are consistently truncated in the expansion in the large scale. The UV divergences can then be absorbed by replacing all fields and parameters of the theory by their renormalised counterparts, multiplied by scheme-dependent renormalisation constants. A novel feature of the SMEFT not present in the SM is the mixing of different operators. In particular, the renormalisation of SMEFT operators requires the introduction of non-physical operators, which also need to be taken into account during the renormalisation process~\cite{Collins_1984, Dixon:1974ss, Kluberg-Stern:1975ebk, Joglekar:1975nu, Deans:1978wn, hep-ph/9409454, 2302.00022,2004.03576}.

The main purpose of this work is to take further steps in the on-going effort to renormalise the dimension-six SMEFT at two loops. We focus in this paper on a subset of operators, and we compute the 
 two-loop QCD corrections to the quark and gluon self-energies, including the contribution from a specific subset of SMEFT dimension-six operators. Our results allow us to extract the contributions from these operators to the field and mass renormalisation constants for quarks and gluons in the modified minimal-subtraction ($\overline{\text{MS}}$) scheme. These renormalisation constants are an important ingredient for two-loop corrections to SMEFT processes involving quarks and gluons. Moreover, we perform our computations using the background-field method \cite{Abbott:1980hw, Abbott:1983zw}, which allows us to extract the running of the strong coupling constant at two-loop order directly from the self-energies. Consequently, we determine the contribution of specific SMEFT operators to the RG running of the top-quark mass and the QCD $\beta$-function.

This work is organised as follows. In Section \ref{sec: QCD-SMEFT} we set up our framework, by defining the operators we consider in this study. In Section \ref{sec: Renormalisation}  we discuss the renormalisation procedure and the redundant operators and we also review the one-loop results. In Section \ref{sec: Two-loop-procedure} we discuss our methodology of the two-loop computation, with the extraction of the counterterms. In Section \ref{sec: running-couplings}, we present the results for the anomalous dimensions for the strong coupling and the top quark mass at two loops. We finally conclude in Section \ref{sec: Conclusions}. We include appendices where we collect the SM results for the renormalsation constants and we review different conventions of normalising the SMEFT operators.

\section{The QCD sector of the SMEFT}
\label{sec: QCD-SMEFT}
In the context of the SMEFT with only dimension-six operators, the Lagrangian takes the form
\begin{equation}
     \mathcal{L}_{\text{SMEFT}}= \mathcal{L}_{\text{SM}}+\sum_i \frac{c_i}{\Lambda^2} \mathcal{O}_i\,,
\end{equation}
where $\Lambda$ is the scale of new physics and $c_i$ are the dimensionless Wilson coefficients. It is well known that not all operators $\mathcal{O}_i$ are independent, but they can be related, e.g., via field redefinitions and/or equations of motion.
In many cases, in particular for the SMEFT at dimension-six, it is well-known how to construct a set of independent gauge-invariant operators~\cite{1008.4884}. Efforts to extract bases of operators at higher dimensions resulted in a complete basis at dimension eight~\cite{2005.00008,2005.00059} and a generalisation of the basis construction at any mass dimension \cite{2201.04639, Harlander:2023psl, Harlander:2023ozs}. However, even if one focuses on dimension six and reverts back to a restricted setting with a simplified flavour structure, the number of independent operators can be large. In applications, one therefore typically limits the analysis to a subset of operators that contribute to a given set of observables.

In view of phenomenological studies at hadron colliders, we focus on the subsector of the dimension-six SMEFT Lagrangian that involves only particles charged under the QCD gauge group SU$(3)$, i.e., quarks or gluons. In the following, we give a description of the operators that we want to consider. First, note that we can classify dimension-six operators according to the number of fermion fields involved. 
If we work in the basis of~\cite{1008.4884} and we restrict to CP-conserving operators, there is a single independent operator without any fermions:
\begin{equation}
\begin{split}
    &  \mathcal{O}_G = f^{a b c} G_{\mu}^{a,\nu} G_{\nu}^{b,\rho} G_{\rho}^{c,\mu}\,, 
\end{split}
\label{eq : 0F-operators}
\end{equation}
where $f^{abc}$ are the SU$(3)$ structure constants and $G_{\mu\nu}^a$ is the gluon field-strength tensor,
\begin{equation}
    G_{\mu \nu}^a = \partial_\mu G_\nu^a -\partial_\nu G_\mu^a - g_s f^{abc}G_\mu^b G_\nu^c\,,
    \label{eq: gluon field-strength}
\end{equation}and $g_s$ is the strong coupling constant.
The operator in Eq.~\eqref{eq : 0F-operators} modifies the three- and four-gluon vertices and introduces five- and six-point vertices.

Next, let us discuss dimension-six operators involving gluon fields and exactly two quark fields. It is easy to see that if we work in the basis of operators of~\cite{1008.4884}, then there are no such operators before electroweak symmetry breaking. Indeed, due to the chiral nature of the SU$(2)_L$ gauge symmetry, we need to include a Higgs doublet to form a gauge-invariant operator. Let us therefore consider the chromomagnetic dipole operators:
\begin{equation}
\begin{split}
 &  \mathcal{O}_{uG}^{ij} = i \Bar{Q}_i \, \tau^{\mu \nu}  T^a  \Tilde{\phi}\,u_j \, G_{\mu \nu}^a + \text{h.c.}\,,\\
 &  \mathcal{O}_{dG}^{ij} = i \Bar{Q}_i \, \tau^{\mu \nu}  T^a  \phi\,d_j \, G_{\mu \nu}^a + \text{h.c.}\,,
\end{split}
\label{eq : 02F-operators}
\end{equation}
where $Q$ and $u/d$ respectively denote the SU$(2)_L$ quark doublet and the right-handed up/down-quark singlet (with generation indices $i,j$) and $\phi$ is the Higgs doublet, with $\tilde{\phi}=i\sigma^2 \phi^*$. Moreover, $T^a$ denote the generators of the fundamental representation of the Lie algebra of SU$(3)$, and $\tau^{\mu\nu} = \frac{1}{2} \comm{\gamma^\mu} {\gamma^\nu}$. 
After electroweak symmetry breaking, these operators 
modify the quark-gluon  interactions and also generate quark-gluon-Higgs vertices. Hence, these operators
reduce to the class of operators we want to consider when the Higgs field is replaced by its vacuum expectation value (vev). 
It is easy to see that a physical observable consistently truncated to $\mathcal{O}\!\left(\tfrac{1}{\Lambda^2}\right)$ only receives contributions from the interference of the SMEFT and the SM. The chromomagnetic operators $\mathcal{O}_{qG}^{ij}$, $q\in\{u,d\}$, flip the chirality of the quark, and since in pure-QCD the quark flavour is conserved, the only source of chirality-flip is the quark-mass term. In the following, we only consider the top quark to be massive, and the remaining $(n_f-1)$ quarks are massless. It is then easy to see that only the operator $\mathcal{O}_{uG}^{33}$ contributes to $\mathcal{O}\left(\tfrac{1}{\Lambda^2}\right)$. We therefore only focus on this particular instance of the chromomagnetic operator. 

Finally, let us comment on four-quark operators. They have the generic structures
\begin{equation}
    \Bar{\Psi}_i \gamma^\mu  \Psi_j  \Bar{\Psi}_k \gamma_\mu   \Psi_l\,,\qquad\Bar{\Psi}_i \gamma^\mu  T^a \Psi_j  \Bar{\Psi}_k \gamma_\mu  T^a \Psi_l\,,
\end{equation}
with $\Psi_i \in \{Q_i, u_i, d_i\}$. Operators of this form have been widely studied in the context of higher-order computations, due to their relevance to flavour-sensitive observables~\cite{Buras:1989xd,Buras:1991jm,hep-ph/9211304,arXiv:hep-ph/0005183, hep-ph/9304257, hep-ph/9311357, hep-ph/9512380,  hep-ph/0312090, hep-ph/0411071, hep-ph/0512066, hep-ph/0612329, 1704.06639, 2402.00249, 2501.08384, 2208.10513,2211.09144, 2211.01379, 2306.16449, 2401.16904,2310.13051}. When computing higher-order corrections, these operators introduce new challenges when working in dimensional regularisation not encountered for zero- and two-fermion operators. First, these operators naturally involve chiral fermions, which raises the question of how to treat $\gamma^5$ in dimensional regularisation in $D=4-2\varepsilon$ dimensions~\cite{Breitenlohner:1977hr, Buras:1989xd}. Second, while we can easily obtain a basis of four-quark operators in four dimensions, this basis is no longer valid in $D$ dimensions, because some identities among spinors used to extract this basis (e.g., Fierz identities) only hold in four dimensions, and one also needs to consider so-called \emph{evanescent operators} that vanish in four dimensions. Due to this increased complexity when dealing with four-fermion operators in dimensional regularisation, 
we postpone the study of these operators and only focus on the dimension-six operators involving either zero or two quark fields.

To sum up, in the remainder of this paper we will focus on the following Lagrangian
\begin{equation}
    \mathcal{L}_{\textrm{QCD},6}= \mathcal{L}_{\text{QCD}} +   \frac{c_{tG}^0}{\Lambda^2}\frac{v}{\sqrt{2}}  \,\mathcal{O}_{tG} + \frac{c_G^0}{\Lambda^2} \,\mathcal{O}_G  \, ,
    \label{eq: Lagrangian}
\end{equation}
where $\mathcal{L}_{\textrm{QCD}}$ denotes the pure-QCD Lagrangian with gauge group SU$(N)$ and $n_f$ quarks, $v$ is the vev, and we defined the dimension-five and six operators
\begin{equation}\begin{split}\label{eq:op_def}
\mathcal{O}_{tG} &\,=i \Bar{t}^0 \tau^{\mu \nu} T^a t^0\, G_{\mu \nu}^{0,a}\, , \\
\mathcal{O}_G &\,= f^{a b c} G_{\mu}^{0,a,\nu} G_{\nu}^{0,b,\rho} G_{\rho}^{0,c,\mu}\,.
\end{split}\end{equation} 
 We denote the bare quantities by a superscript 0, and the bare top mass will be denoted by $m^0$ (recall that all other quarks are considered massless). 
We also assume the Lagrangian to be CP-conserving, which in particular forces $c_{tG}$ to be real. 

\section{Renormalisation}
\label{sec: Renormalisation}
The main goal of our paper is to compute the field and mass renormalisation counterterms at two loops in the $\overline{\textrm{MS}}$-scheme for the Lagrangian defined in Eq.~\eqref{eq: Lagrangian}. Since Eq.~\eqref{eq: Lagrangian} derives from the full SMEFT Lagrangian, these renormalisation constants describe a subset of the ultraviolet (UV) divergences that arise in the full SMEFT if we focus on QCD interactions only. We start by discussing the structure of field and mass renormalisation constants in general. An important aspect of the renormalisation at two-loop order is the handling of subdivergences (both UV and infrared (IR)), and we discuss these issues in Section~\ref{sec:off-shell}.

\subsection{UV renormalisation of two-point functions}
The field and mass renormalisation constants of a QFT can be extracted from the computation of the two-point Green's functions of the theory to a given loop order. More specifically, we may consider the off-shell two-point Green's function for a quark or gluon in dimensional regularisation in $D=4-2\varepsilon$ dimensions. If we work off-shell, the Green's function is IR finite~\cite{Poggio:1976qr,PhysRevD.14.2123}, and all poles in the dimensional regulator $\varepsilon$ are of UV origin. 

The renormalisation is performed by replacing the bare fields and parameters in Eq.~\eqref{eq: Lagrangian} by renormalised ones and including relevant unphysical operators for the renormalisation of redundant structures, as discussed in detail in Section \ref{sec:off-shell}. 
Recall that we denote bare quantities by a superscript 0, and quantities without a superscript refer to the renormalised ones. We introduce the renormalisation constants
\begin{equation}\begin{split}
    \phi^0 &\,= \sqrt{Z_\phi} \,\phi\,,\qquad \phi\in\{G, q, t\},\\
     g_s^0 &\,= Z_{g_s} \,g_s\,, \\
          m^0 &\,= Z_{m}\, m\,, \\
          \xi^0 &\,= Z_\xi \, \xi \, ,\\
      c_i^0 &\,= Z_{ij}\, c_j\,,\qquad  \ c_i \in \{c_{tG}, c_G\}\,,
    \label{eq: general_renormalisation}
    \end{split}
\end{equation}
where $q$ runs over the $(n_f-1)$ light-quark fields. The renormalisation factors of the fields and parameters are given as expansions in perturbation theory, both in $\alpha_s = \tfrac{g_s^2}{4 \pi}$ and $\tfrac{1}{\Lambda}$, 
\begin{equation}
    Z_i = 1 + \sum_{L\geq 1} \qty(\frac{\alpha_s}{4 \pi})^{L-1}\left[ \frac{\alpha_s}{4 \pi}\,\delta Z_i^{(L), 4} + \frac{g_s}{16 \pi^2}\,\frac{c_{tG}}{\Lambda^2} \,\delta Z_{i,tG}^{(L), 6} + \frac{g_s}{16 \pi^2}\,\frac{c_{G}}{\Lambda^2}\, \delta Z_{i,G}^{(L), 6} \right]+ \mathcal{O}\left(\frac{1}{\Lambda^{3}}\right)\,,
    \label{eq: renormalisation_factors}
\end{equation}
where $i \in \{G, t, q, g_s,m, \xi\}$, and $\alpha_s$ is the renormalised strong coupling constant in the $\overline{\textrm{MS}}$-scheme.\footnote{Note that in the $\overline{\textrm{MS}}$-scheme the renormalised coupling depends on the renormalisation scale $\mu$.}  The factors $ \delta Z_i^{(L), 4} $ correspond to the pure-QCD (dimension four) renormalisation factors and they are known up to five loops~\cite{Tarrach:1980up,Tarasov:1982plg,Larin:1993tq,Chetyrkin:1997dh,Vermaseren:1997fq,Baikov:2014qja,Luthe:2016xec,Baikov:2017ujl,Luthe:2017ttc}.  The remaining terms $\delta Z_{i,tG}^{(L), 6} $ and $\delta Z_{i,G}^{(L), 6} $ account for the contributions from dimension-six operators.

Let us make a comment about our conventions for the normalisation of the effective operators in Eq.~\eqref{eq : 02F-operators}, specifically how it affects the counting of the powers of the strong coupling. 
As defined in Eq.~\eqref{eq : 02F-operators}, the operators $\mathcal{O}_G$ and $\mathcal{O}_{tG}$ generate contributions to the pure gluon and top-gluon vertices that are not proportional to powers of $ g_s $. This is reflected in the perturbative expansion of the Green's functions and the renormalisation constants, cf.~Eq.~\eqref{eq: renormalisation_factors}. In particular, the dimension-six contributions contain one less overall power of the strong coupling compared to the SM quantity computed at the same order. For example, in Eq.~\eqref{eq: renormalisation_factors}, only a single power of $ g_s $ multiplies the EFT contribution at one-loop, with additional powers of $ \alpha_s $ starting to appear only at two loops. This issue is avoided, for example, in the conventions of \texttt{SMEFT@NLO}~\cite{2008.11743},\footnote{A SMEFT model used for phenomenological analysis in NLO QCD.} by modifying the definitions of the operators in Eq.~\eqref{eq : 02F-operators} to include an extra factor of \( g_s \), ensuring that the EFT contribution matches the QCD one in terms of powers of \( \alpha_s \). We discuss how our conventions are related to those of \texttt{SMEFT@NLO} in Appendix~\ref{sec: SMEFTatNLO}, where we also show the corresponding results.

Finally, let us briefly comment on the renormalisation of the Wilson coefficients. We have
\begin{equation}
    Z_{ij} = \delta_{ij} + \sum_{L\geq 1}\qty(\frac{\alpha_s}{4 \pi})^L \delta Z_{ij}^{(L)}+\mathcal{O}\left(\frac{1}{\Lambda^2}\right)\,,
    \label{eq: WCs renormalisation}
\end{equation}
and the $\delta Z_{ij}^{(L)}$ are generally non-diagonal, leading to mixing among effective operators. The complete one-loop anomalous dimension of the dimension-six SMEFT for a specific basis of operators was obtained in~\cite{1308.2627,1310.4838,1312.2014}. 
From the knowledge of the matrix of one-loop counterterms $\delta Z^{(1)}_{ij}$, it is possible to extract the one-loop anomalous dimensions.

When renormalising the two-point Green's functions at two loops, one needs to consistently subtract one-loop subdivergences.
During this process, insertions of higher-dimensional operators generally produce unphysical tensor structures that must be considered, particularly in a fully off-shell framework. We therefore revisit the one-loop renormalisation in the next subsection.

\subsection{Off-shell renormalisation}
\label{sec:off-shell}
As is well known \cite{Collins_1984, Dixon:1974ss, Kluberg-Stern:1975ebk, Joglekar:1975nu, Deans:1978wn}, renormalising a general gauge theory requires more than just its initial ingredients. The operators in Eq.~\eqref{eq : 02F-operators} are gauge-invariant, \textit{physical} operators (class I) that contribute to the S-matrix. Upon renormalisation, they mix either with other class I operators or with \textit{unphysical} operators (class II), which do not contribute to the S-matrix. Class II operators are further divided into gauge-invariant operators that vanish via equations of motion  (class II.a) and gauge-variant operators, BRST-exact operators (class II.b). In our approach, the gauge-variant operators of class II.b arise because, when quantising the theory, gauge invariance is explicitly broken, and only BRST symmetry is retained. 
Operators of class II do not contribute to physical processes, because they either vanish due to the equations of motion, or they are BRST-exact, and BRST-exact operators do not alter physical matrix elements. They do, however, contribute to off-shell Green's functions. 

Let us discuss in a bit more detail what this implies for our Lagrangian in Eq.~\eqref{eq: Lagrangian}. We start by recalling the equations of motions for the quark fields and the gluon field-strength tensor,
\begin{equation}\begin{split}
    (i \slashed{D} - m) t&\,=0 + \mathcal{O}\left(\tfrac{1}{\Lambda^2}\right)\,, \\ 
    i \slashed{D}q&\,=0+ \mathcal{O}\left(\tfrac{1}{\Lambda^2}\right)\,, \\ 
    D^\mu G_{\mu \nu}^a&\,=g_s \sum_{\psi\in \{t,q\}}\Bar{\psi}\gamma_\nu T^a \psi+ \mathcal{O}\left(\tfrac{1}{\Lambda^2}\right)\,,
    \label{eq: EOMs}
    \end{split}
\end{equation}
where $\slashed{D} = \gamma^{\mu}D_{\mu}$ and we introduced the covariant derivative 
\begin{equation}
    D_\mu = \partial_\mu+ig_sG_\mu^aT^a_R\,,
    \label{eq: covariant-derivative}
\end{equation} 
with $T_R^a$ the generator of the representation $R$ of SU$(N)$, i.e., $T_R^a=T^a$ is the covariant derivative if it acts on a quark and $(T_R^a)_{bc} = -if^{abc}$ if it acts on a gluon field. Note that, if we work at linear order in the dimension-six SMEFT, it is sufficient to only consider the SM equations of motion. 

In order to renormalise the Green's functions off-shell, we need to add to the Lagrangian in Eq.~\eqref{eq: Lagrangian} operators of class II,
\begin{equation}
    \mathcal{L} =\mathcal{L}_{\textrm{QCD},6} +  \sum_{i}\frac{c_i^{\textrm{II},0}}{\Lambda^2} \mathcal{O}_i^{\textrm{II},0}\,,
\end{equation}
where the sum runs over all bare operators $\mathcal{O}_i^{\textrm{II},0}$ of class II and up to dimension six that we can construct out of quark and gluon fields (we again follow the convention that bare quantities are denoted by a superscript 0). In the renormalisation of two-loop diagrams, the insertion of these operators as counterterm vertices cancels the subdivergences of the two-loop amplitude arising from one-loop subgraphs. We will list the operators that are relevant for this paper below. While these operators do not contribute to physical observables, they need to be taken into account when renormalising the theory. In particular, operators of class II may only mix among themselves under renormalisation, making their contribution unphysical to all orders in perturbation theory. However, mixings of operators of class I into operators of class II need to be taken into account. Specifically, while class I coefficients are renormalised according to Eqs.~\eqref{eq: general_renormalisation} and \eqref{eq: WCs renormalisation}, the Wilson coefficients $c_i^{\textrm{II},0}$ get renormalised according to
\begin{equation}
    c_i^{\textrm{II},0} = Z_{i,G}^{\textrm{I}}\, c_G+Z_{i,tG}^{\textrm{I}}\, c_{tG}+ \sum_{j}Z_{ij}^{\textrm{II}}\, c_j^{\textrm{II}}\,.
\end{equation}
In practice, we only need to consider $Z_{ij}^{\textrm{I}}$, $j\in\{G,tG\}$, which are determined through the renormalisation of off-shell Green's functions with insertions of physical operators. In general, they admit the perturbative expansion
\begin{equation}
    Z_{ij}^{\textrm{I}} = \sum_{L \geq 1} \qty(\frac{\alpha_s}{4 \pi})^{L-1}\, \frac{g_s}{16 \pi^2}\, \delta Z_{ij}^{\textrm{I},(L)}\,.
    \label{eq: classII-ct}
\end{equation}

We have performed the renormalisation of all relevant Green's functions at one-loop fully off-shell in the $\overline{\textrm{MS}}$-scheme. In addition to the two-point functions, this required the computation of one-loop corrections to the quark-gluon and three-gluon vertices. Renormalisation in a fully off-shell framework does not require IR subtraction, as off-shell Green's functions are inherently IR-finite~\cite{Poggio:1976qr,PhysRevD.14.2123}. We work with the background-field method~\cite{Abbott:1980hw, Abbott:1983zw} and decompose the gluon field $G_\mu^a$ into a background field $B_\mu^a$ and a quantum field $Q_\mu^a$, such that $G_\mu^a = B_\mu^a + Q_\mu^a$. In this context, a specific choice of the gauge-fixing Lagrangian, the background-field gauge, ensures that explicit gauge invariance is retained for the background field. As a result, class II.b operators involving the background field do not appear in the computation. However, at the two-loop level, subgraphs that involve corrections to the quantum field may still require counterterm insertions from class II.b operators involving the latter~\cite{2203.11181, hep-ph/9409454}. In this framework, the gluon renormalisation factor, $Z_G$, is decomposed into separate renormalisation factors for the quantum and background gluons, denoted as $Z_Q$ and $Z_B$, respectively. Although the renormalisation of Green's functions with external background gluons does not require the determination of $Z_Q$, we compute it nonetheless, as its expression is equivalent to the gluon renormalisation factor in the covariant gauge. 
\sloppy
The one-loop Feynman diagrams relevant for our computation are generated with $\texttt{FeynArts}$~\cite{hep-ph/0012260}, using a model built in $\texttt{FeynRules}$~\cite{1310.1921, 2010.15852}. The analytic expressions for the one-loop diagrams are manipulated with the packages $\texttt{FeynCalc}$~\cite{2312.14089} and $\texttt{FeynHelpers}$~\cite{1611.06793}. We can express all Feynman diagrams in terms of one-loop integrals whose pole structure is known. This allows us to identify both class II operators generated through mixing as well as the renormalisation factors. Our findings are summarised in Table~\ref{tab: one-loop ct}. The first column lists the two physical operators $\mathcal{O}_G$ and $\mathcal{O}_{tG}$, while the second column contains all the operators generated at the one-loop level, along with their respective class and counterterm. We also use the notation $\{.\}_Q$ to indicate that the operator involves only quantum fields and is irrelevant to renormalising the background field $B_\mu^a$. We have chosen our basis of class II operators in such a way that vanishings due to the equations of motion in Eq.~\eqref{eq: EOMs} are manifest. We note that for the triple-gluon operator $\mathcal{O}_G$, operators containing four-fermion structures are generated through redundant operators, potentially leading to mixing into these structures. However, this effect cancels when these operators are combined with their respective counterterms, reflecting the well-known fact that no mixing occurs with four-fermion operators at one-loop~\cite{1312.2014}. In Table~\ref{tab: one-loop ct} we also include the class II operators $\mathcal{O}_{\partial^3t}$ and $\mathcal{O}_{\partial^3q}$, which are not generated at one-loop, but will become relevant starting at two loops.  Moreover, the EFT contribution to the QCD renormalisation factors at one-loop reads
\begin{align}
     \delta Z_{t,tG}^{(1),6}&\,=\frac{v}{\sqrt{2}} \frac{6 \, C_F m}{\varepsilon}\,,  \notag\\   
     \delta Z_{m,tG}^{(1),6}&\, = \frac{v}{\sqrt{2}} \frac{12 \, C_F m}{\varepsilon}\,, \notag\\  
     \delta Z_{Q,tG}^{(1),6}&\, = \delta Z_{\xi,tG}^{(1),6} \,= \delta Z_{B,tG}^{(1),6} \,=\frac{v}{\sqrt{2}}  \frac{8 m }{\varepsilon}\,, \notag\\ 
     \delta Z_{g_s,tG}^{(1),6}&\,= - \frac{v}{\sqrt{2}} \frac{4 m }{\varepsilon}  \,.
    \label{eq: one-loop-CT-EFT}
\end{align}
When writing our results, we use the usual expressions for the quadratic Casimir operators in the fundamental and adjoint representations of SU$(N)$,
\begin{equation}
\CA=N \textrm{~~~and~~~} \CF=\frac{N^2-1}{2N}\,.
\end{equation} The results for the renormalisation of the top field, top mass and strong coupling constant agree with the one-loop results presented in \cite{1503.08841,1607.05330}.
\fussy

\renewcommand{\arraystretch}{2.0}
\setlength{\extrarowheight}{1pt}
\begin{table}[h]
\small
    \centering
    \begin{tabular}{|c|c|c|c|}
        \hline
          \makecell{Physical \\ Operator}  &  Operators generated at loop-level  &   Class &  One-loop counterterm  \\ 
        \hline
        \hline
        \multirow{3.3}{*}{\centering $\mathcal{O}_{tG} 
        $}& $\mathcal{O}_{tG} = i \Bar{t} \tau^{\mu \nu} T^a t G_{\mu \nu}^a$ & I & $\dfrac{30 C_F -23 C_A + 2 n_f}{6 \varepsilon} $\\ & $ \mathcal{O}_{\partial^2t} = \Bar{t}(i \slashed{D} - m)^2t$ & II.a &  $\dfrac{v}{\sqrt{2}} \dfrac{6 C_F }{\varepsilon} $\\  \hfill
        & $ \left\{g_s(\Bar{t}\slashed{G}(i \slashed{D} - m)t - \Bar{t}(i \overleftarrow{\slashed{D}}+ m)\slashed{G} t)\right\}_Q$ & II.b & $\dfrac{v}{\sqrt{2}}\dfrac{3 C_A }{4 \varepsilon} $\\ [1.5ex]
        \hline \hline
        \multirow{8.4}{*}{\centering $\mathcal{O}_{G} 
        $}& $\mathcal{O}_{tG}=  i \Bar{t} \tau^{\mu \nu} T^a t G_{\mu \nu}^a$ & I & $\dfrac{3 m\, C_A}{\sqrt{2}\, v\, \varepsilon}$\\ & $\mathcal{O}_{G} = f^{a b c} G_{\mu}^{a,\nu} G_{\nu}^{b,\rho} G_{\rho}^{c,\mu}$ & I & $\dfrac{C_A+ 2 n_f}{2 \varepsilon}$\\  & $\mathcal{O}_{\partial^3t} = \Bar{t}(i \slashed{D}-m)^3t$ & II.a & $0$\\
  & $\mathcal{O}_{\partial^3q} = \Bar{q}(i \slashed{D})^3\,q$ & II.a & $0$\\      
        & $\mathcal{O}_{\partial^2G} =(D^\mu G_{\mu \nu}^a - g_s \Bar{\psi_i} \gamma_\nu T^a \psi_i)^2$ & II.a & $\dfrac{3 C_A}{\varepsilon}$\\ 
         & $\mathcal{O}_{\psi\psi\partial G}=g_s \Bar{\psi_j} \gamma^\nu T^a \psi_j(D^\mu G_{\mu \nu}^a - g_s \Bar{\psi_i} \gamma_\nu T^a \psi_i)$ & II.a & $\dfrac{3 C_A}{\varepsilon}$\\ 
           & $\left\{g_s f^{a b c}(D^\mu G_{\mu \nu}^a - g_s \Bar{\psi_i} \gamma_\nu T^a \psi_i)G_{\nu \rho}^b G_{\rho}^c\right\}_Q$ & II.b & $\dfrac{3 C_A}{2 \varepsilon}$\\ 
          & $\left\{g_s f^{a b c}(D^\mu G_{\mu \nu}^a - g_s \Bar{\psi_i} \gamma_\nu T^a \psi_i) \partial^\nu \Bar{c}^b c^c\right\}_Q$ & II.b& $\dfrac{3 C_A}{2 \varepsilon}$\\ [1.5ex]
        \hline
    \end{tabular}
    \caption{Summary of operators of class I and II (second column) generated at one-loop via an insertion of the physical operators $\mathcal{O}_G$ and $\mathcal{O}_{tG}$ (first column) defined in Eq.~\eqref{eq:op_def} with their respective class (third column) and counterterms (fourth column), and $\psi_i \in \{ t, q\}$. A sum over the repeated flavour index $i$ is understood. }
    \label{tab: one-loop ct}
\end{table}
\normalsize

\section{Two-loop renormalisation of quark and gluon self-energies}
\label{sec: Two-loop-procedure}

In this section we present the main results of this paper, namely the two-loop renormalisation constants for the quark and gluon two-point Green's functions. 
We start by generating the relevant two-loop Feynman diagrams for the top, light-quark and gluon self-energies with $\texttt{FeynArts}$. We then project the expressions for the Feynman diagrams onto scalar form factors. Specifically, the quark self-energy can be decomposed into two independent spinor structures
\begin{equation}
    \hat\Sigma(p^2,m) = \Sigma_V(p^2,m) \slashed{p} + \Sigma_S(p^2,m) m \,,
\end{equation}
with $p$ the off-shell momentum of the external quark. The scalar functions can be extracted with the following projectors as
\begin{equation}
    \Sigma_V(p^2,m) = \frac{1}{4 p^2} \Tr\left[\slashed{p}\hat\Sigma(p^2,m)\right]\,, \qquad \Sigma_S(p^2,m) = \frac{1}{4 m} \Tr\left[\hat\Sigma(p^2,m)\right]\,.
\end{equation}
The gluon self-energy can be written as
\begin{equation}
    \hat \Pi_{\mu \nu}(p^2) = \Pi(p^2)(p^2 g_{\mu \nu} - p_\mu p_\nu)\,,
\end{equation}
where we extract the scalar form factor $\Pi(p^2)$ by applying the projector
\begin{equation}
    \Pi(p^2) = P_{\mu \nu}  \hat\Pi_{\mu \nu}(p^2)\,, \qquad  P_{\mu\nu}=\frac{1}{p^4(D-1)}(p^2 g_{\mu \nu} - p_\mu p_\nu)\,.
\end{equation}

\begin{figure}[h]
\centering
    \begin{subfigure}{0.3\textwidth}
    \centering
    \includegraphics[width=1\linewidth]{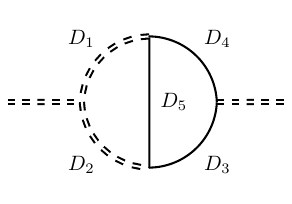}
    \end{subfigure}
\quad
    \begin{subfigure}{0.3\textwidth}
    \centering
    \includegraphics[width=1\linewidth]{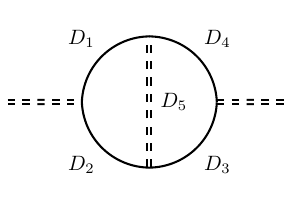}
    \end{subfigure}
\quad
    \begin{subfigure}{0.3\textwidth}
    \centering
    \includegraphics[width=1\linewidth]{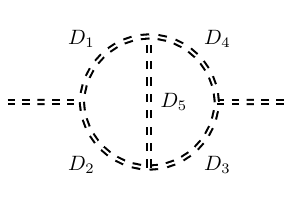}
    \end{subfigure}
    \caption{\label{fig:G_topos}Two-loop topologies associated to the gluon self-energy. Massive propagators are represented by solid lines, while the massless ones appear as double dashed segments. Following the convention in Table \ref{tab: gluon_families}, we have families $G_A$, $G_B$ and $G_C$ on the left, centre and right, respectively. }
\end{figure}

At the end of this procedure, we have reduced the problem to the computation of the scalar form factors. 
The Dirac and color algebra has been carried out via a mixed use of $\texttt{FORM}$~\cite{math-ph/0010025} and $\texttt{FeynCalc}$. 
The scalar form factors can then be written as linear combinations of scalar Feynman integrals organised in terms of integral families of the form
\begin{equation}
I^{F}_{n_1 n_2 n_3 n_4 n_5} = e^{2\varepsilon \gamma}\int \prod_{j=1}^2 \frac{d^Dk_j}{i \pi^{D/2}}\frac{1}{D_1^{n_1}D_2^{n_2}D_3^{n_3}D_4^{n_4}D_5^{n_5}},
\end{equation}
where $F$ labels the independent families determined by the propagators $D_i$, $\gamma=-\Gamma'(1)$ is the Euler-Mascheroni constant, and the $n_i$ are integers.
\begin{figure}[h]
\centering
    \begin{subfigure}{0.3\textwidth}
    \centering
    \includegraphics[width=1\linewidth]{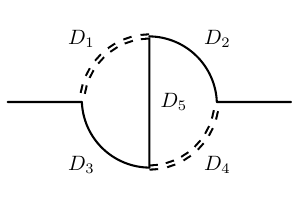}
    \end{subfigure}
\qquad\quad
    \begin{subfigure}{0.3\textwidth}
    \centering
    \includegraphics[width=1\linewidth]{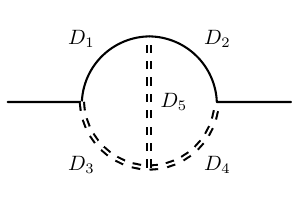}
    \end{subfigure}
    \caption{\label{fig:T_topos}Two-loop topologies associated to the top-quark self-energy. Massive propagators are represented by solid lines, while the massless ones appear as double dashed segments. Following the convention in Table \ref{tab: top_families}, we have families $T_A$ and $T_B$ on the left and right, respectively. }
\end{figure}
We identify two integral families $T_A$ and  $T_B$ that contribute to the top self-energy at two loops, defined by the propagators in Table \ref{tab: top_families} (see Fig.~\ref{fig:T_topos}). For the gluon self-energy we have three families $G_A$, $G_B$ and $G_C,$ defined in Table \ref{tab: gluon_families} (see Fig.~\ref{fig:G_topos}). The self-energy for the light quarks can be expressed in terms of the same families as in the gluon case.
\renewcommand{\arraystretch}{1.2}
\begin{table}[ht]
    \centering
    \begin{subtable}{0.3\linewidth}
    \centering
        \begin{tabular}{c|c}
          $T_A$ &  $T_B$ \\ \hline
          $k_1^2$ & $k_1^2-m^2$ \\
          $k_2^2-m^2$ & $k_2^2-m^2$ \\
          $(k_1-p)^2-m^2$ & $(k_1-p)^2$\\
          $(k_2-p)^2$ & $(k_2-p)^2$\\
          $(k_1-k_2)^2-m^2$ & $(k_1-k_2)^2$
    \end{tabular}
    \caption{}
    \label{tab: top_families}
    \end{subtable}
    \begin{subtable}{0.65\linewidth}
    \centering
    \hspace{1cm}
        \begin{tabular}{c|c|c}
           $G_A$ &  $G_B$&  $G_C$\\ \hline
          $k_1^2$ & $k_1^2-m^2$ & $k_1^2$\\
          $(k_1+p)^2$ & $(k_1+p)^2-m^2$ & $(k_1+p)^2$\\
          $(k_2+p)^2-m^2$ & $(k_2+p)^2-m^2$ & $(k_2+p)^2$\\
          $k_2^2-m^2$ & $k_2^2-m^2$ & $k_2^2$\\
          $(k_1-k_2)^2-m^2$ & $(k_1-k_2)^2$ & $(k_1-k_2)^2$
    \end{tabular}
    \caption{}
    \label{tab: gluon_families}
    \end{subtable}
    \caption{Definition for the set of integral families required for  calculation of (a) top-quark and (b) gluon and light-quark self-energies at two loops. }
    \label{tab: Integral_families}
\end{table} 

Every integral from any of the five families in Table~\ref{tab: Integral_families} can be expressed as a linear combination of master integrals (MIs) specific to that family. 
We perform the reduction to MIs using integration-by-parts (IBP)~\cite{Chetyrkin:1979bj,Tkachov:1981wb, hep-ph/0102033} identities using  \texttt{Kira}~\cite{1705.05610,2008.06494}. For the families contributing to the top-quark self-energy, we find 8 MIs, all belonging to $T_A$,
\begin{equation}
    I^{T_A}_{01100},\, I^{T_A}_{01101},\,I^{T_A}_{01102},\,I^{T_A}_{10011},\,I^{T_A}_{10012},\,I^{T_A}_{11100},\,I^{T_A}_{11110},\,I^{T_A}_{11111}\,,
\end{equation}
while for the gluon case we find 10 MIs
\begin{equation}
    I^{G_A}_{00101},\, I^{G_A}_{00111},\,I^{G_A}_{11010},\, I^{G_A}_{11110},
    \,  I^{G_A}_{10202}, \, I^{G_A}_{20201}, \, I^{G_A}_{11111}, \,I^{G_B}_{11110},\,I^{G_C}_{10101},\,I^{G_C}_{11110}\,.
\end{equation}
The expressions for the MIs expanded in $\varepsilon$ are well known and were taken from the literature \cite{hep-ph/9803493, hep-ph/0008287, hep-ph/0611236,Adams:2016xah,Adams:2017ejb}. Note that some of the master integrals involve functions of elliptic type~\cite{Adams:2016xah,Adams:2017ejb}. Those functions, however, only contribute to the finite parts, and are irrelevant for the renormalisation. The amplitudes have been handled in a \texttt{Mathematica} code with the help of the packages \texttt{PolyLogTools} \cite{1904.07279} and \texttt{HypExp} \cite{0708.2443}. We note that the integral families encountered in our SMEFT computation are identical to those encountered in a pure-QCD context. This should not come as a surprise, because the effect of the dimension-six operators is to modify the numerators of the Feynman diagrams, while the propagator structures remain unchanged compared to the SM.

Since we are working in the $\overline{\textrm{MS}}$-scheme, we can focus on the poles of the self-energies. We have already mentioned that the off-shell self-energy is free of IR poles, and so all poles are of UV origin. However, the UV poles appearing in the two-loop self-energy do not immediately translate into the two-loop renormalisation constants. Indeed, starting from two loops, we need to take into account the appearance of one-loop subdivergences, and they need to be subtracted consistently to isolate the two-loop renormalisation constants. This subtraction is encoded into the $R$-operation~\cite{Zimmermann:1969jj,Bogoliubov:1957gp,Hepp:1966eg}. At this stage, the computation of the renormalisation constants in the EFT shows some novel features not present in the SM. Indeed, in the EFT, we need to take into account contributions from one-loop subdivergences due to the insertion of operators of class II. We now illustrate this new feature with some explicit examples.

\begin{figure}[!h]
    \centering
    \includegraphics[width=0.65\linewidth]{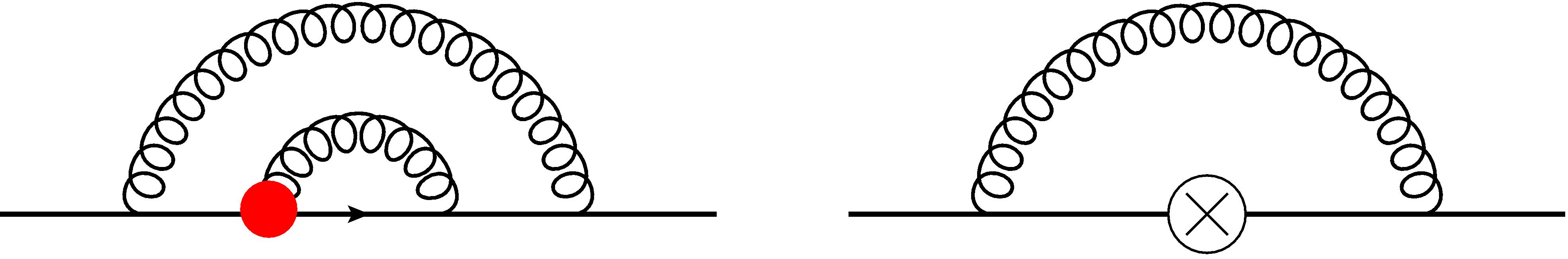}
    \caption{Example Feynman diagrams for the quark self-energy at two-loop (left) and one-loop with the insertion of a counterterm (right) where we also insert class II.a operators when needed, due to the insertion of $\mathcal{O}_{tG}$, in red.}
    \label{fig: top self-energy diag}
\end{figure}

As a first example, consider the two-loop diagram in Fig.~\ref{fig: top self-energy diag}. It contains the one-loop subgraph corresponding to a correction to the top-quark propagator. The subdivergences arising from this diagram will be cancelled by the one-loop diagram with an insertion of the counterterm to the top-quark line. In this case, also the class II.a operator $\mathcal{O}_{\partial^2t}$ is inserted (see Table \ref{tab: one-loop ct}), as it was needed to renormalise the off-shell two-point function at one-loop. 
Proceeding this way, we can subtract the subdivergences for both the quark and gluon self-energies using the operators listed in Table \ref{tab: one-loop ct}. As expected from the background-field method, gauge-variant operators of class II.b which involve the background field were not needed for the renormalisation. 

\begin{figure}[h]
    \centering
    \includegraphics[width=0.6\linewidth]{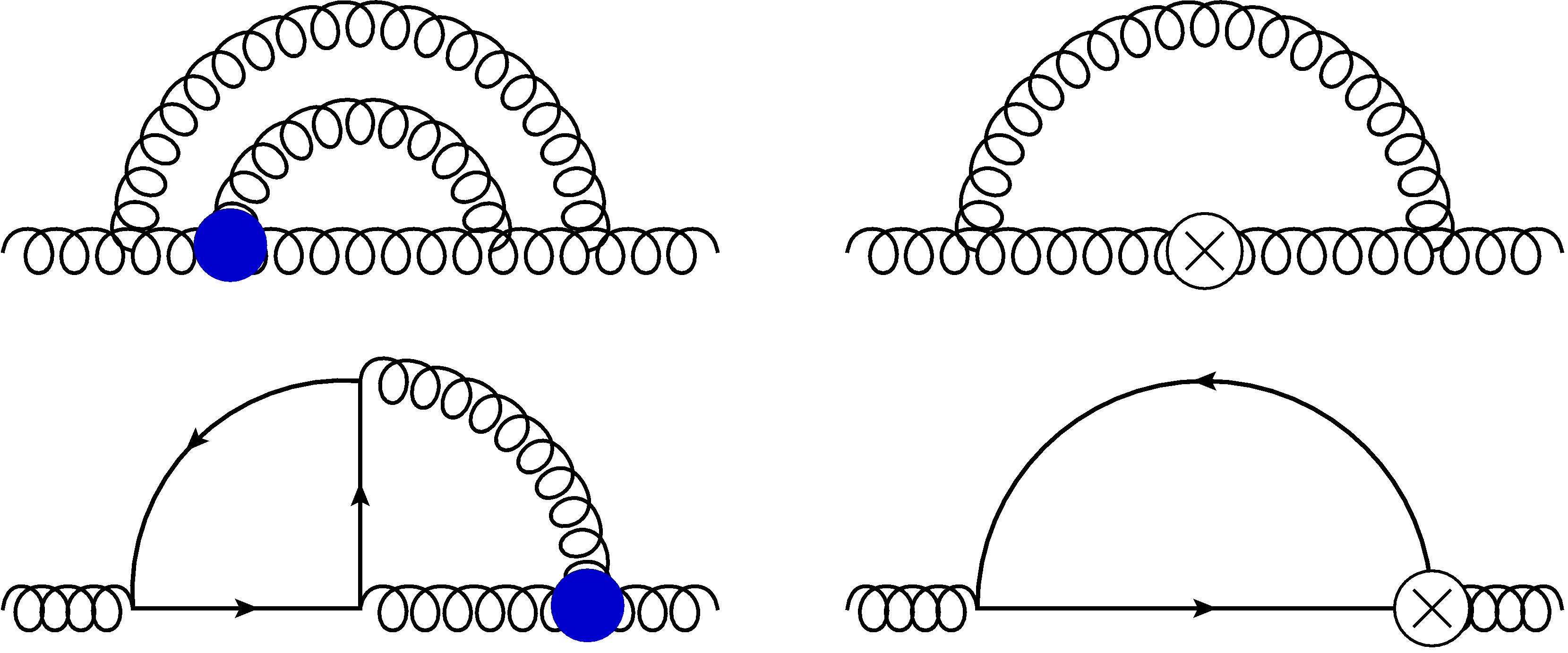}
    \caption{Examples of Feynman diagrams for the gluon self-energy at two-loop (left) and one-loop with the insertion of counterterm (right) with subgraphs that correct the gluon propagator (top panel) and $\Bar{\psi} \psi G$ vertex (bottom panel) due to an insertion of $O_G$, in blue.}
    \label{fig: Gluon self-energy diag}
\end{figure}

As another example, in Fig.~\ref{fig: Gluon self-energy diag} we show a two-loop Feynman diagram that arises through the insertion of the class I operator $\mathcal{O}_G$. The subtraction of the one-loop subdivergence requires us to consider the 
 dimension-six operator  
$ \mathcal{O}_{\partial^2G} $ (see Table~\ref{tab: one-loop ct}). Similarly, the operator  
$ \mathcal{O}_{\psi\psi\partial G} $  
accounted for the subgraphs related to the $ \Bar{\psi} \psi G $ vertex correction.  

\begin{figure}[h]
    \centering
    \includegraphics[width=\linewidth]{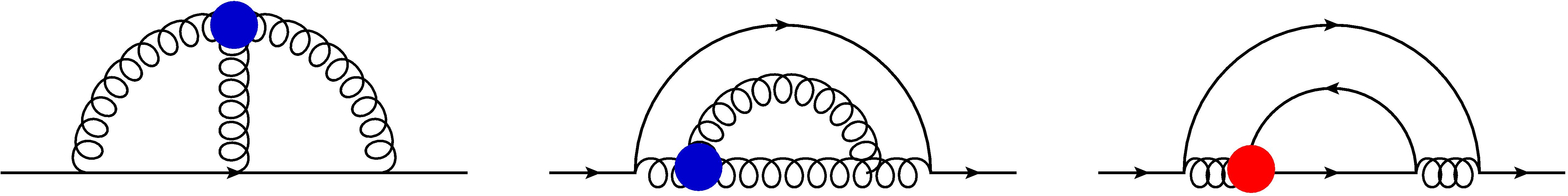}
    \caption{Example Feynman diagrams for quark self-energy at two loops, showing the first insertion of the class I operator $\mathcal{O}_G$ (blue) and the first insertion of the class I operator $\mathcal{O}_{tG}$ (red) in the light-quark self-energy, induced by a top-quark loop subgraph.}
    \label{fig: quarkself}
\end{figure}

As a final example, we consider the Feynman diagrams contributing to the two-loop quark self-energy. 
We note that, while at one-loop the class I operator $\mathcal{O}_G$ does not contribute to the quark self-energy due to the absence of diagrams involving three-gluon vertices, at the two-loop level we have the first contribution. 
Being a purely dimension-six operator, $\mathcal{O}_G$ produces a higher-order structure in the derivative expansion with respect to the dipole operator, which instead acts as a dimension-five operator. Specifically, the structure $\slashed{p} p^2$ is generated, which can be renormalised by the additional dimension-six class II.a operators $\mathcal{O}_{\partial^3 t} $ and $\mathcal{O}_{\partial^3 q}$ (see Table~\ref{tab: one-loop ct})
respectively for the top and light-quark fields.
Similarly, while the dipole operator would not contribute to the light-quark self-energy at one-loop, it enters the computation at two loops due to the subgraphs with the top-quark loop, effectively contributing to the light-quark field renormalisation factor. Examples of these diagrams are shown in Fig. \ref{fig: quarkself}.

We also observe that to cancel the non-localities of the pole, it was necessary to renormalise the gauge-fixing parameter $\xi$, with the corresponding renormalisation factor $Z_\xi$ determined to have the same expression as $Z_Q$.

Let us now discuss our results.
After expanding all MIs in $\varepsilon$ and summing the two-loop diagrams with the one-loop counterterm insertions, the non-local pole structure cancels. This is then used to determine the two-loop renormalisation factors in Eq.~\eqref{eq: renormalisation_factors}. Note that we have also computed the pure QCD diagrams, and we reproduce the known results for the corresponding two-loop renormalisation constants (see Appendix~\ref{sec: two-loop CTs}). The contributions from dimension-six operators are:

\begin{align}
\delta Z_{t, tG}^{(2),6}= & \, - \frac{\CF m \,v}{4 \sqrt{2}} \bigg[\frac{1}{ \varepsilon^2}\Big(24(\xi-1)\CF + 77 \CA - 8 n_f \Big)\notag\\
&\,  + \frac{1}{6 \varepsilon}\Big(8(93 \CF + 17 n_f +18)-931 \CA \Big)\bigg]\,,\notag \\  \delta Z_{t, G}^{(2),6}= & \,  \frac{3 m^2 \CF \CA}{2} \bigg[\frac{3}{\varepsilon^2} + \frac{1}{\varepsilon}\bigg], \notag\\
\delta Z_{q, tG}^{(2),6}= & \, - \frac{6 \CF m \, v}{\sqrt{2} \varepsilon}, \notag\\
\delta Z_{q, G}^{(2),6}= & \, 0, \notag\\
\delta Z_{m, tG}^{(2),6}= & \, -  \CF \frac{m \,v}{\sqrt{2}} \bigg[\frac{2}{\varepsilon^2}\Big( 12 \CF + 17 \CA - 2(n_f+3)\Big)  + \frac{1}{3 \varepsilon}\Big(60 \CF -175 \CA +10 n_f+114 \Big)\bigg], \notag\\ \delta Z_{m, G}^{(2),6}= &\, m^2 \CF \CA \bigg[\frac{9}{ \varepsilon^2} - \frac{7}{4 \varepsilon}\bigg],\notag\\
\delta Z_{Q,tG}^{(2),6}= & \,  \frac{m \,v}{\sqrt{2}} \bigg[\frac{2}{\varepsilon^2}\Big( 4 \CF - (2 \xi+7) \CA\Big) + \frac{1}{\varepsilon}\Big(24 \CF + 13 \CA \Big)\bigg],\notag\\  \delta Z_{Q,G}^{(2),6}= & \,  3 m^2 \CA\bigg[\frac{2}{ \varepsilon^2} - \frac{23}{\varepsilon}\bigg], \notag\\
\delta Z_{B,tG}^{(2),6}= & \,  \frac{4 m \,v}{\sqrt{2}} \bigg[\frac{2}{\varepsilon^2}\Big( \CF- \CA\Big) + \frac{1}{\varepsilon}\Big(6 \CF + 5 \CA \Big)\bigg],\notag\\  \delta Z_{B,G}^{(2),6}= &\, 3 m^2 \CA\bigg[\frac{2}{ \varepsilon^2} - \frac{23}{\varepsilon}\bigg]\,.
\label{eq: two-loop dim6 counterterms}
\end{align}
In addition to the counterterms for the QCD fields and parameters, we determine those for the following class II operators that affect the two-point functions,
\begin{equation}
\begin{aligned}
    & \mathcal{O}_{\partial^2 t}\,, \quad \mathcal{O}_{\partial^3 t}\,, \quad \mathcal{O}_{\partial^3 q} \,, \quad
    \mathcal{O}_{\partial^2 B} = \left\{\mathcal{O}_{\partial^2 G}\right\}_B, \ \ \mathcal{O}_{\partial^2 Q} = \left\{\mathcal{O}_{\partial^2 G}\right\}_Q\,.
\end{aligned}
\label{eq: two-loop EOM operators}
\end{equation}

We note that the operator $\mathcal{O}_{\partial^2 G}$ requires different counterterms depending on whether it involves quantum or background fields.\footnote{Even though the presence of this operator suggests possible mixing into four-fermion operators, no definitive statement can be made at this stage, as contributions from three- and four-point Green's functions are still missing.} Following Eq.~\eqref{eq: classII-ct}, we find
\begin{align}
\delta Z_{\partial^2t, tG}^{\textrm{I},(2)}= & \, - \frac{\CF v}{4 \sqrt{2}} \bigg[ \frac{1}{\varepsilon^2}\Big(12(\xi-5)\CF - (3\xi-68)\CA - 8 n_f \Big) \notag\\  
& - \frac{1}{6 \varepsilon}\Big(24(\xi-16)\CF + (649-6 \xi)\CA-88 n_f\Big)\bigg] \,,\notag\\\delta Z_{\partial^2t, G}^{\textrm{I},(2)}= & \, \frac{9 m \CF \CA}{2 \varepsilon^2} \,, \notag\\
\delta Z_{\partial^3t, G}^{\textrm{I},(2)}=  &\, \, \delta Z_{\partial^3q, G}^{\textrm{I},(2)} = \,  \frac{\CF \CA}{4 \varepsilon}\,,\notag\\
\delta Z_{\partial^2B, G}^{\textrm{I},(2)}= & \, - \frac{\CA}{4} \bigg[\frac{1}{2 \varepsilon^2}(3 \CA \xi+7 \CA-16 n_f)-\frac{1}{3 \varepsilon}((9 \xi+47) \CA-40 n_f)\bigg]\,, \notag\\ 
\delta Z_{\partial^2Q, G}^{\textrm{I},(2)}= & \, - \frac{\CA}{4} \bigg[\frac{1}{2 \varepsilon^2}(3 \CA \xi-16 n_f)-\frac{1}{3 \varepsilon}((9 \xi+62) \CA-40 n_f)\bigg]\,,
\label{eq: two-loop EOM counterterms}
\end{align}
where we work in $R_{\xi}$ gauge, and $\xi$ is the corresponding gauge-fixing parameter.

\section{The running of the top-mass and the strong coupling constant}
\label{sec: running-couplings}

It is well known that renormalisation constants give access to the anomalous dimensions governing the renormalisation group running. In particular, in the $\overline{\textrm{MS}}$-scheme the top mass runs, and the running is governed by the renormalisation group equation
\begin{equation}
    \frac{\dd  m(\mu)}{\dd\! \log \mu} = \gamma_m = \gamma_m^{(4)} + \gamma_m^{(6)}+ \mathcal{O}\left(\tfrac{1}{\Lambda^3}\right)\,,
\end{equation}
where the superscripts $(4)$ and $(6)$ refer again to the contribution from pure-QCD and dimension-six operators. The pure-QCD contribution is known through five loops~\cite{Tarrach:1980up,Tarasov:1982plg,Larin:1993tq,Chetyrkin:1997dh,Vermaseren:1997fq,Baikov:2014qja,Luthe:2016xec,Baikov:2017ujl,Luthe:2017ttc}, and we reproduce those results up to two loops. We note that the mass counterterm is $\xi$-independent. Focusing on the contribution from dimension-six operators, we  find
\begin{equation}
\begin{split}
    \gamma_m^{(6)} = & \, m^2 \ctG \frac{g_s}{16 \pi^2} 4 \CF \bigg[6 - \frac{\alpha_s}{4 \pi}\,\frac{1}{3}(60 \CF -175 \CA +10 n_f + 114)+ \mathcal{O}\qty(\frac{\alpha_s}{4 \pi})^2 \bigg] \\ & - m^3  \cG \frac{g_s}{16 \pi^2}\left[\frac{\alpha_s}{4 \pi}\, 7 \CF \CA+ \mathcal{O}\qty(\frac{\alpha_s}{4 \pi})^2\right]\,.
\end{split}
\end{equation} The one-loop result agrees with that presented in \cite{1503.08841}.
  
Our computation also allows us to extract the running of the strong coupling constant. Indeed, since the computation was performed in the background-field gauge, the strong coupling renormalisation factor $Z_{g_s}$ can be extracted without requiring the computation of two-loop vertex corrections, because we have the relation \cite{Abbott:1980hw}:
\begin{equation}
Z_{g_s} \sqrt{Z_B} = 1\,.
\end{equation}
Using Eqs.~\eqref{eq: one-loop-CT-EFT} and~\eqref{eq: two-loop dim6 counterterms}, we find
\begin{equation}\begin{split}
\delta Z_{g_s, tG}^{(2),6}= & \, - \frac{2 m \,v}{\sqrt{2}}\bigg[\frac{1}{\varepsilon^2}\Big( 2\CF- 13\CA + 2 n_f\Big) + \frac{1}{\varepsilon}\Big(6 \CF + 5 \CA \Big)\bigg]\,,\\ \delta Z_{g_s, G}^{(2),6}= & \, - 3 m^2 \CA \bigg[\frac{1}{ \varepsilon^2} - \frac{23}{2 \varepsilon}\bigg]\,.
\end{split}\end{equation}
The knowledge of the renormalisation constant $Z_{g_s}$ can immediately be turned into the QCD $\beta$-function that governs the running of the strong coupling constant in the $\overline{\textrm{MS}}$-scheme:
\begin{equation}
    \frac{\dd  \alpha_s(\mu)}{\dd\! \log \mu} = \beta=\beta^{(4)} + \beta^{(6)} + \mathcal{O}\left(\tfrac{1}{\Lambda^3}\right)\,.
\end{equation}
The pure-QCD contribution $\beta^{(4)}$ is known through five loops~\cite{vanRitbergen:1997va,hep-ph/9912391,Czakon:2004bu,Baikov:2016tgj,1709.08541}, and we reproduce the one- and two-loop QCD results.
The contribution to the running of the strong coupling due to the operators $\mathcal{O}_G$ and $\mathcal{O}_{tG}$ is
\begin{equation}
\begin{split}
    \beta^{(6)} = & \,-16 \alpha_s  m \ctG \frac{g_s}{16 \pi^2}  \bigg[ 1 + \frac{\alpha_s}{4 \pi} (6 \CF + 5 \CA) + \mathcal{O}\qty(\frac{\alpha_s}{4 \pi})^2\bigg] \\ & + 276 \alpha_s m^2 \cG \frac{g_s}{16 \pi^2} \left[\frac{\alpha_s}{4 \pi}\,\CA + \mathcal{O}\qty(\frac{\alpha_s}{4 \pi})^2\right]\,.
\end{split}
\end{equation}
We observe that the contribution of the triple-gluon operator to the running of the strong coupling is proportional to the mass of the top quark. Consequently, in a model containing only bosons or massless fermions, this contribution would be zero, aligning our findings with those reported in Ref.~\cite{2410.07320}.

\section{Conclusions}
\label{sec: Conclusions}
In this work, we have computed the two-loop QCD corrections to the quark and gluon two-point functions, including contributions from a subset of SMEFT dimension-six operators, constructed from gluon and quark fields. We assume that only the top quark is massive, and the remaining $n_f-1$ quarks are massless. 

The computation was performed using the standard methodology to compute multiloop amplitudes. We have generated all diagrams up to two loops that contribute to the quark and gluon self-energies with at most one insertion of a higher-dimensional operator. The resulting expressions are projected onto scalar form factors, which can be cast in the form of a linear combination of scalar integrals from different families. These families of integrals have been computed in the literature. In particular, for each of them a set of master integrals is known. After inserting the analytic expression for the master integrals, we can extract the UV poles of the self-energies up to two loops.

We perform the renormalisation of the self-energies in the $\overline{\textrm{MS}}$-scheme, and we work with the background-field method. This allows us to extract the contributions from the dimension-six operators to the renormalisation constants for the quark and gluon fields, as well as for the top mass and the strong coupling, which is the main result of our work.
A main feature of our computation is that the renormalisation requires the introduction of additional unphysical, redundant operators. These operators, inserted into one-loop counterterm diagrams, were necessary to cancel subdivergences arising from one-loop subgraphs with dimension-six operator insertions. 

Our computation is a further step towards the renormalisation of the dimension-six SMEFT at two loops, and our results are a crucial ingredient for two-loop computations of processes involving quarks and gluons in this theory. For the future, it would be interesting if we could extend our approach to extract renormalisation constant for other SMEFT quantities, including other renormalisations schemes, e.g., the on-shell scheme.

\paragraph{Note added:} While our computations were being completed, we became aware of~\cite{2412.13251}, which has some partial overlap with our results. In particular, the contributions from the chromomagnetic operator $\mathcal{O}_{tG}$ to the renormalisation constants was presented there, and we have checked that we independently reproduce those results. The results for the gluon operator $\mathcal{O}_{G}$ are new and presented here for the first time.

\section*{Acknowledgements}
GV thanks Hesham El Faham, Victor Miralles and Rudi Rahn for useful discussions, and CD acknowledges discussions with Ani Venkata about the IR-finiteness of off-shell Green's functions. GV and EV are supported by the European Research Council (ERC) under the European
Union’s Horizon 2020 research and innovation programme (Grant agreement No. 949451) and by a Royal
Society University Research Fellowship through grant URF/R1/201553. The work of AV is supported by the DFG project 499573813 “EFTools”.

\appendix
\section{Renormalisation factors in QCD up to two loops}
\label{sec: two-loop CTs}
The pure-QCD renormalisation constants are known up to five loops~\cite{Tarrach:1980up,Tarasov:1982plg,Larin:1993tq,Chetyrkin:1997dh,Vermaseren:1997fq,Baikov:2014qja,Luthe:2016xec,Baikov:2017ujl,Luthe:2017ttc,vanRitbergen:1997va,hep-ph/9912391,Czakon:2004bu,Baikov:2016tgj,1709.08541}. As a side product of the renormalisation of the SMEFT contributions, we have reproduced the pure renormalisation constants of a SU($N$) pure QCD using a general $\xi$-gauge up to two loops in the $\overline{\textrm{MS}}$-scheme, and we summarise them here. At one-loop we have
\begin{equation}
\begin{aligned}
  &  \delta Z_\psi^{(1),4} = -\frac{C_F \xi}{\varepsilon}\,, \qquad \delta Z_m^{(1),4} = - \frac{3 C_F}{\varepsilon}\,, \qquad \delta Z_Q^{(1),4} = \frac{(13-3\xi)C_A-4 n_f}{6 \varepsilon} \,,\\
  &  \delta Z_B^{(1),4}  =\frac{11C_A-2 n_f}{3 \varepsilon}\,, 
  \qquad \delta Z_{g_s}^{(1),4} = -\frac{11 C_A-2 n_f}{6 \varepsilon}\,,
  \label{eq: one-loop-CT-QCD}
\end{aligned} 
\end{equation}
At two loops we find
\begin{equation}
\begin{aligned}
\delta Z_\psi^{(2),4}= & \, \frac{\CF \xi}{4 \varepsilon^2}\Big(2 \xi \CF +(\xi+3) \CA \Big) + \frac{\CF}{8 \varepsilon}\Big(6 \CF - (\xi(\xi+8)+25)\CA +4 n_f\Big)\,,\\ 
\delta Z_m^{(2),4}=& \, \frac{\CF}{2 \varepsilon^2}\Big(9 \CF +11 \CA - 2n_f\Big) - \frac{\CF}{12 \varepsilon}\Big(9 \CF +97 \CA-10 n_f\Big)\,, \\  \delta Z_Q^{(2),4}= & \,\frac{\CA}{24 \varepsilon^2}\Big(2 \xi +3 \Big)\Big((3\xi-13)\CA + 4 n_f\Big) \\ & - \frac{1}{16 \varepsilon}\Big((\xi(2\xi+11)-59)\CA^2+ 16 n_f \CF + 20 n_f \CA \Big)\,,\\
 \delta Z_B^{(2),4} = & \, \frac{1}{3 \varepsilon} \Big(17 \CA^2-3 n_f \CF-5 n_f \CA  \Big)\,, \\  \delta Z_{g_s}^{(2),4} = & \, \frac{1}{24 \varepsilon^2} \Big(11 \CA - 2 n_f  \Big)^2-\frac{1}{6 \varepsilon} \Big(17 \CA^2 - 3 n_f \CF - 5 n_f \CA  \Big)\,. 
\end{aligned}
\end{equation}
The mass anomalous dimension and the QCD $\beta$-function read
\begin{align}
    \gamma_m^{(4)} = & \, - m C_F \bigg[\qty(\frac{\alpha_s}{4 \pi}) 6  + \qty(\frac{\alpha_s}{4 \pi})^2 \frac{1}{3}(9 \CF + 97 \CA - 10 n_f)+\mathcal{O}\qty(\frac{\alpha_s}{4 \pi})^3\bigg]\,,\\
  \nonumber  \beta^{(4)} = & \,- 2 \alpha_s \bigg[\qty(\frac{\alpha_s}{4 \pi})\frac{1}{3}(11 \CA - 2 n_f) + \qty(\frac{\alpha_s}{4 \pi})^2\frac{2}{3}(17 \CA^2-3 n_f \CF - 5 n_f \CA)+\mathcal{O}\qty(\frac{\alpha_s}{4 \pi})^3\bigg]\,.
\end{align}

\section{Comparison with SMEFT@NLO conventions}
\normalsize
\label{sec: SMEFTatNLO}

The results obtained in this work were computed following the Warsaw basis conventions \cite{1008.4884}, where the relevant dimension-six operators are defined as in Eq. \eqref{eq : 02F-operators}, with the conventions set in Eqs. \eqref{eq: gluon field-strength} and \eqref{eq: covariant-derivative}. As explained in Section \ref{sec: Renormalisation}, these conventions mean that the amplitude with an insertion from the dimension-six operators  at a given loop order has one less overall power of $g_s$ compared to the corresponding QCD amplitude.

In the conventions of \texttt{SMEFT@NLO} \cite{2008.11743}, a \texttt{UFO} \cite{Degrande:2011ua,Darme:2023jdn} model used for phenomenological analyses at NLO in QCD, this imbalance is avoided by introducing a factor of the strong coupling constant in the definition of the operators. To be precise, in the conventions of \texttt{SMEFT@NLO} we have
\begin{equation}
    \mathcal{O}_{tG} = i g \, \Bar{Q} \, \tau^{\mu \nu}  T^a  \Tilde{\phi}\,t \, G_{\mu \nu}^a + \text{h.c.}\,, \qquad \mathcal{O}_G = g f^{a b c} G_{\mu \nu}^a G_{\nu \rho}^b G_{\rho \mu}^c\,,
\end{equation}
where $ \Bar{Q} $ denotes the SU$(2)_L $ quark doublet of the third generation, and $ t $ represents the right-handed top-quark field. The difference in the notation for the strong coupling constant, which we denoted with $g$ instead of $g_s$ as in the rest of this paper, arises from an additional difference  in the conventions of \texttt{SMEFT@NLO}. Specifically, in \texttt{SMEFT@NLO} we have
\begin{equation}
    G_{\mu \nu}^a = \partial_\mu G_\nu^a -\partial_\nu G_\mu^a + g f^{abc}G_\mu^b G_\nu^c\,, \qquad   D_\mu \psi = \partial_\mu\psi-ig G_\mu^aT^a\psi\,.
\end{equation}
In this context, the expansion of the renormalisation factors in Eq.~\eqref{eq: renormalisation_factors} reads
\begin{equation}
    Z_i = 1 + \sum_{L\geq 1} \qty(\frac{\alpha}{4 \pi})^L \delta Z_i^{(L), 4} + \sum_{L \geq 1} \qty(\frac{\alpha}{4 \pi})^{L} \delta Z_{i}^{(L), 6} + \mathcal{O}(\Lambda^{-3})\,,
    \label{eq: renormalisation_factors_SMEFTatNLO}
\end{equation}
so that both the dimension-four and dimension-six contributions have the same power of $\alpha= \tfrac{g^2}{4\pi}$. However, due to the change in conventions, the insertions from $ \mathcal{O}_{tG} $ and $ \mathcal{O}_G $ acquire an overall minus sign, causing the sign of the counterterms $ \delta Z_i^{(L),6} $ shown in Eq.~\eqref{eq: two-loop dim6 counterterms} to be flipped. The same applies to the counterterms for the class II operators in Table~\ref{tab: one-loop ct} and Eq.~\eqref{eq: two-loop EOM counterterms}.  

Additionally, the renormalisation factors for the Wilson coefficients are modified, as an extra factor of $ Z_g = Z_{g_s} $ multiplies the renormalised vertex. The renormalisation factors $ \Bar{Z}_{c_{tG}} $ and $ \Bar{Z}_{c_G} $ in the \texttt{SMEFT@NLO} conventions can be straightforwardly obtained, as they are related to those in the Warsaw conventions by $ \Bar{Z}_{i} = Z_i / Z_{g_s} $.  

To conclude, in the \texttt{SMEFT@NLO} conventions, the anomalous dimension for the top-quark mass and the QCD $\beta$-function in the presence of the dimension-6 operators are given by the following expressions:
\begin{align*}
    \gamma_m^{(6)} = & \,- 4 m^2 \ctG \CF  \bigg[\qty(\frac{\alpha_s}{4 \pi}) 6 - \qty(\frac{\alpha_s}{4 \pi})^2 \frac{1}{3}(60 \CF -175 \CA +10 n_f + 114) \bigg] \\ & + m^3  \cG \qty(\frac{\alpha_s}{4 \pi})^2 7 \CF \CA+\mathcal{O}\qty(\frac{\alpha_s}{4 \pi})^3\,, \\
    \beta^{(6)} = & \, 16 \alpha_s m \ctG  \bigg[ \qty(\frac{\alpha_s}{4 \pi}) + \qty(\frac{\alpha_s}{4 \pi})^2  (6 \CF + 5 \CA)\bigg] \\ & - 276 \alpha_s m^2 \cG \qty(\frac{\alpha_s}{4 \pi})^2   \CA+\mathcal{O}\qty(\frac{\alpha_s}{4 \pi})^3\,. \tag{B.4}
\end{align*}

\bibliographystyle{JHEP.bst}
\bibliography{main.bib}

\end{document}